\documentclass{aa}  

\usepackage{graphicx}
\usepackage{txfonts}
\usepackage{hyperref}
\usepackage{xcolor}
\usepackage[normalem]{ulem}
\usepackage{amsmath}	% Advanced maths commands
\usepackage{amssymb}	% Extra maths symbols
\usepackage{listings}
\usepackage{multirow}
\usepackage{rotating}

\begin{document}

   \title{The effect of winds on atmospheric layers of red supergiants}
%   \title{Spectroscopy of bright BA-type stars in NGC\,300 with MUSE}
	
	%\subtitle{I. 1D modelling for AH Sco, UY Sct and KW Sgr}

   \subtitle{I. Modelling for interferometric observations}

\author{G. Gonz\'alez-Tor\`a
          \inst{1,2}
          \and
          M. Wittkowski\inst{1}
          \and
          B. Davies\inst{2}
          \and
          B. Plez\inst{3}
          \and
          K. Kravchenko\inst{4}}

   \institute{European Southern Observatory (ESO),
   Karl Schwarzschildstrasse 2, 85748 Garching bei München, Germany\\
        \email{Gemma.GonzaleziTora@eso.org}
   \and
        Astrophysics Research Institute, Liverpool John Moores
        University, 146 Brownlow Hill, Liverpool L3 5RF, United Kingdom\
    \and
       LUPM, Universit\'e de Montpellier, CNRS, 34095 Montpellier, France
       \and
       Max Planck Institute for extraterrestrial Physics, Giessenbachstrasse 1, 85748 Garching bei München, Germany}    

   \date{Received \today; accepted \today}

% \abstract{}{}{}{}{} 
% 5 {} token are mandatory
 
  \abstract
  % context heading (optional)
  % {} leave it empty if necessary  
   {Red supergiants (RSGs) are evolved massive stars in a stage preceding core-collapse supernova. The physical processes that trigger mass loss in their atmospheres are still not fully understood, and remain one of the key questions in stellar astrophysics. Based on  observations of $\alpha$ Ori, a new semi-empirical method to add a wind to hydrostatic model atmospheres of RSGs was recently developed. This method can reproduce many of the static molecular shell (or 'MOLsphere') spectral features. %without relying on dusty models. 
   }
  % aims heading (mandatory)
   {We use this method of adding a semi-empirical wind to a {\sc MARCS} model atmosphere to compute synthetic observables, comparing the model to spatially resolved interferometric observations. %We use both a simple radiative equilibrium model and a semi-empirical model including a chromospheric temperature inversion.%We use this semi-empirical method to study the extended atmospheres of RSGs for interferometric data, studying the effect of a stellar wind for both a simple radiative equilibrium model and a semi-empirical model based on observations of the atmosphere of $\alpha$ Ori. 
  We present a case study to model published interferometric data of HD 95687 and V602~Car obtained with the AMBER instrument at the Very Large Telescope Interferometer (VLTI). 
   }
  % methods heading (mandatory)
   {We compute model intensities with respect to the line of sight angle ($\mu$) for different mass-loss rates, spectra and visibilities using the radiative transfer code {\sc Turbospectrum}. We can convolve the models to match the different spectral resolutions of the VLTI instruments, studying a wavelength range of $1.8-5\,\mathrm{\mu m}$ corresponding to the $K$, $L$ and $M$-bands for GRAVITY and MATISSE data. The model spectra and squared visibility amplitudes are compared with the published VLTI/AMBER data. }
  % results heading (mandatory)
   {The synthetic visibilities reproduce observed drops in the CO, SiO, and water layers that are not shown in visibilities based on {\sc MARCS} models alone. For the case studies, we find that adding a wind on to the MARCS model %$\dot{M}=10^{-5.5}$ $M_{\odot}/\mathrm{yr}$ (HD~95687) and $\dot{M}=10^{-5}$ $M_{\odot}/\mathrm{yr}$ (V602~Car)
   with simple radiative equilibrium dramatically improves the agreement with the squared visibility amplitudes as well as the spectra, the fit being even better  when applying a steeper density profile than predicted from previous studies. Our results reproduce observed extended atmospheres up to several stellar radii.
   }
   % Although the mass loss is extremely high compared to the typical ranges for RSGs ($\sim \dot{M}=10^{-6}$ $M_{\odot}/\mathrm{yr}$) \citep{2020MNRAS.492.5994B}, we especulate that this high mass loss range could be localized in a region of the star, as suggested by \citet{2022arXiv220107818H,2022MNRAS.510..383A}. }
  % conclusions heading (optional), leave it empty if necessary 
   {This paper shows the potential of our model to describe extended atmospheres in RSGs. It can reproduce the shapes of the spectra and visibilities with better accuracy in the CO and water lines than previous models. The method can be extended to other wavelength bands for both spectroscopic and interferometric observations. We provide temperature and density stratifications that succeed for the first time in reproducing observed interferometric properties of red supergiant atmospheres.}


   \maketitle

%-------------------------------------------------------------------

\section{Introduction}\label{sec:int}
%What are Red supergiants and how their mass loss affects their life.
Stellar winds impact the lives of massive stars and can change their evolutionary path in the Hertzprung-Russell diagram (HRD). These mass-loss events become important as the star leaves the main sequence phase \citep{1986ARA&A..24..329C}. For this reason, one of the key stages for mass-loss is the red supergiant (RSG) phase, where the massive star is in a stage preceding core-collapse supernova. These mass-loss events occur in the extended atmospheres of RSGs, whose extensions go up to several stellar radii. Beyond that, the temperature is low enough to produce a dusty shell. However, the observed extensions are not at all reproduced by current dynamic model atmospheres which include pulsation and convection \citep{2015A&A...575A..50A}. 

As a consequence, the mechanism that triggers mass-loss in the extended atmospheres of RSGs is still poorly understood. This is not the case for their low- and intermediate-mass counterparts (Miras), whose mass-loss processes can be explained by pulsation and dust driven winds alone \citep{1979ApJ...227..220W,1988ApJ...329..299B,2018A&ARv..26....1H}. There have been several attempts to explain the mechanism of stellar winds in RSGs \citep[e.g.,][]{2007A&A...469..671J,1986ARA&A..24..329C,2000ARA&A..38..613K}, but there is still no consensus. Recent work by \citet{2021A&A...646A.180K} studied the effect of turbulent atmospheric pressure in initiating and determining the mass-loss rates of RSGs, finding promising results. However, further work is needed to unambiguously determine the dynamical processes that trigger massive stellar wind events in spatially extended atmospheres.
 %In short, the stellar models known to date are missing a physical mechanism to fully explain massive stellar wind events on spatially extended atmospheres. 
%Studies of mass loss in IR and far-IR, little studies in the optical to near-IR. What did Ben's paper show.

Hence, there have been few studies that explore the mass loss effect in cool massive stars, all of them focusing on their spectra. Most explore the mid- and far-IR region, where the dust component is dominant \citep[e.g.,][]{2009A&A...506.1277G,2018MNRAS.475...55B,2006A&A...456..549D,2010A&A...523A..18D}. Therefore, the models rely solely on dust modelling, such as DUSTY \citep{1997MNRAS.287..799I,1999ascl.soft11001I} or RADMC3D \citep{2012ascl.soft02015D}. Recently, \citet{2021MNRAS.508.5757D} explored the extension of the atmospheres close to the stellar surface at radii smaller than the inner dust shells in the optical and near-IR. Adding the influence of a stellar wind in the {\sc MARCS} model atmospheres \citep{2008A&A...486..951G}, \citet{2021MNRAS.508.5757D} expanded the atmosphere up to several stellar radii. Their results naturally explained the presence of mid-IR excess, as well as the mismatch between temperatures derived from the optical and the IR \citep{2005ApJ...628..973L,2013ApJ...767....3D,2021MNRAS.505.4422G}. They also reproduced many of the features obtained by addition of a static molecular shell (or 'MOLsphere'). In short, the work by \citet{2021MNRAS.508.5757D} opened a new window to explore the mass-loss rates of cool massive stars. %that does not need to rely in dusty stellar models. 
%Previous interferometric studies

So far these models have been constrained by comparison to stellar spectra only. The spectral computation shows the flux integrated over the stellar disk and misses the spatially resolved information.  Therefore, if we want a detailed way to study spatially extended stellar atmospheres, we need to use interferometric data. Interferometry uses an array of telescopes to increase the angular resolution of the observations. It is a very powerful tool to study the topography of extended atmospheres in detail, and has been used widely both for RSGs \citep[e.g.,][]{2013A&A...554A..76A,2012A&A...540L..12W,2020A&A...635A.160C,2022A&A...658A.185C} and Miras \citep[e.g.,][]{2018A&A...613L...7W,2020A&A...642A.235K}. %By using interferometry, we have high spatial resolution data of the stellar atmospheres of RSGs. This is an additional information that spectroscopy does not fully provide. 
As a consequence, interferometry represents a stronger test for models.

%What we do in our paper
In this work, we employ the approach by \citet{2021MNRAS.508.5757D} to extend atmospheres for both spectral and interferometric data. We compute the synthetic visibilities of the models for different mass-loss rates ($\dot{M}$) in the wavelength range of the Very Large Telescope Interferometer (VLTI) instruments  ($1.8-5\,\mathrm{\mu m}$), which goes from near-IR to mid-IR. We explore the robustness of the model, and present a case study for VLTI/AMBER \citep{2007A&A...464....1P} data for the RSGs HD~95687 and V602~Car from the sample by \citet{2015A&A...575A..50A}. We have chosen these two targets because they are examples of two stars with different masses and therefore different $\dot{M}$, as well as two distinct luminosities. %We find the best fitting model to these observations and its corresponding $\dot{M}$, comparing with previous studies.
%How the paper is organized.
\medskip

This paper is organised as follows: Section~\ref{sec:analysis} describes the model used for the present study. The theoretical results from our spatially resolved model atmosphere are presented in Section~\ref{sec:results}, followed by the case study for HD~95687 and V602~Car in Section~\ref{sec:case}. %In chapter~\ref{sec:discussion} we explore the implications of the results. 
Lastly, we conclude in Section~\ref{sec:conclusion}. 
%\begin{figure}[t]
%\centering
%\includegraphics[width=1.\linewidth]{}
%\caption{•}
%\label{fig:}
%\end{figure}

%--------------------------------------------------------------------

%\section{Observations and Data Reduction}\label{sec:obs}
%For our case study, we have used published data of the three red supergiants (RSGs) AH Sco, UY Sct, and KW Sgr with the ESO Very Large Telescope Interferometer (VLTI). We used the medium-resolution mode ($R \sim 1500$) in the $K-2.1\,\mathrm{\mu m}$ and $K-2.3\,\mathrm{\mu m}$ bands. The data reduction is explained in detail at \citet{2013A&A...554A..76A}.

%-----------------------------------------------------------------
\section{Methods}\label{sec:analysis}
In absence of models that self-consistently explain winds of RSGs, we add a stellar wind with a constant $\dot{M}$ to an initial {\sc MARCS} model \citep{2008A&A...486..951G}, following the method by \citet{2021MNRAS.508.5757D}. These models are then used to calculate both the synthetic spectra and the squared visibility amplitudes ($|V|^{2}$). This is described in detail in the following sections.

\subsection{Models}\label{sec:models}
\begin{figure*}[t]
\centering
\includegraphics[width=1.\linewidth]{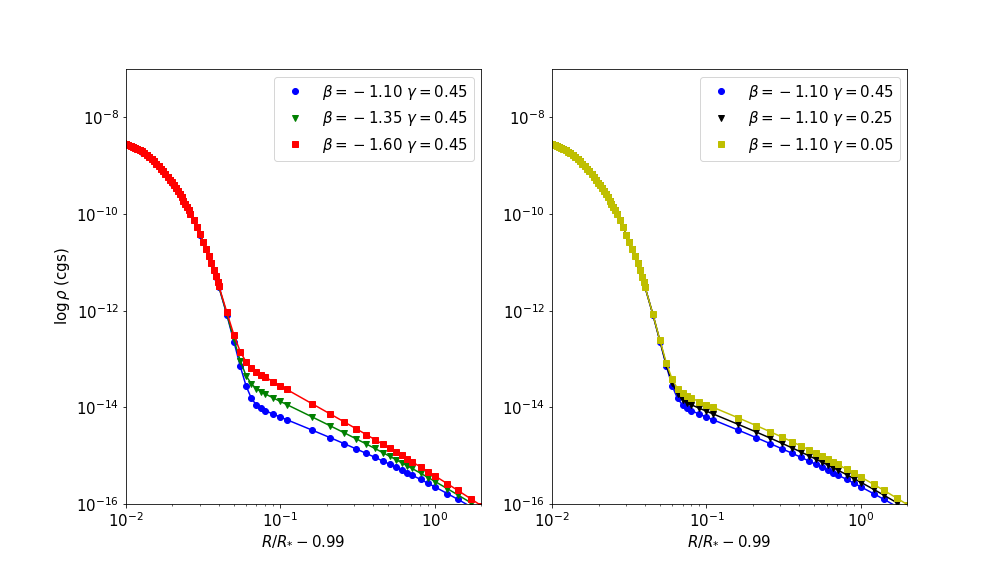}
\caption{\textit{Left: } The different density profiles for $\gamma=0.45$ and variations of $\beta$ with $\beta=-1.10$ (blue dots), $-1.35$ (green triangles) and $-1.60$ (red squares), we see that as we decrease $\beta$, the wind density near $R/R_{\star}-0.99=10^{-1}$ increases and gets steeper. \textit{Right:} The different density profiles for $\beta=-1.10$ and the variations of $\gamma$ with $\gamma=0.45$ (blue dots), $0.25$ (black triangles) and $0.05$ (yellow squares). In this case, as we increase $\gamma$ the slope of the profile remains the same, but the values of the wind density gradually increase. The model used with $\log \mathrm{L}/\mathrm{L}_{\odot}=4.8$,  $T_{\mathrm{eff}}=3500$ K, $\log g=0.0$, $M=15\,M_{\odot}$ and $\dot{M}=10^{-5.5}$ $M_{\odot}/\mathrm{yr}$, corresponds to the stellar parameters of HD~95687.
}
\label{fig:beta-law-grid}
\end{figure*}

We started with a {\sc MARCS} model atmosphere. This code assumes local thermodynamic equilibrium (LTE), hydrostatic equilibrium, and spherical symmetry. We defined a radius grid for the model, allowing to contain a more extended stratification up to $\sim8.5\,R_{\star}$, where $R_{\star}$ is defined as the radius where the Rosseland opacity $\tau_{\mathrm{Ross}}=2/3$. Moreover, for simplicity, we  assumed that:
\begin{itemize}
\item The wind is in LTE. A discussion to this assumption can be found in \citet{2021MNRAS.508.5757D}.
\item The model is 1D, so we assume spherical symmetry. %So far no model exists for RSGs that accounts for possible asymmetries \citep{2018A&ARv..26....1H}. 
%\item The velocity $v$ is zero at all depths, to simplify radiative transfer.
\end{itemize}
To determine the outermost density, we use the mass continuity expression,
\begin{equation}\label{eq:masscont}
\dot{M}=4\pi r^{2}\rho(r)v(r)
\end{equation}
where $\rho$ and $v$ are the density and velocity as a function of the stellar radial coordinate $r$, respectively. The wind density $\rho_{\mathrm{wind}}(r)$ has the shape proposed by \citet{2001ApJ...551.1073H}: 
\begin{equation}\label{eq:betalaw}
\rho_{\mathrm{wind}}=\frac{\rho_{\mathrm{phot.}}}{(R_{\mathrm{max}}/R{\star})^{2}} \left( 1-\left( \frac{0.998}{(R_{\mathrm{max}}/R_{\star})} \right) ^{\gamma} \right) ^{\beta}
\end{equation}
where $R_{\mathrm{max}}$ is the arbitrary outer-most radius of the model, in our case $8.5\,R_{\star}$. The $\beta$ and $\gamma$ parameters define the smoothness of the extended wind region and were initially set in the semi-empirical 1D model of $\alpha$ Ori by \citet{2001ApJ...551.1073H}: $\beta_{\mathrm{Harp}}=-1.10$ and $\gamma_{\mathrm{Harp}}=0.45$. %We investigated the effect of changing these parameters in Section~\ref{sub:beta}.
In Figure~\ref{fig:beta-law-grid} we show what happens when changing the $\gamma$ and $\beta$  parameters that define the density profile. The variations of $\beta$ mostly influence the smoothness of the density profile close to the stellar surface, while the variations of $\gamma$ influence the full density profile to upper or lower values. We will discuss the implications on spectra and interferometric visibility values in Section~\ref{sub:beta}.

%We found the best parameters $\beta_{\mathrm{new}}=-1.60$ and $\gamma_{\mathrm{new}}=0.05$ with means of a $\chi^{2}$ minimization by comparing the model $|V|^{2}$ with the data for our two case studies. They better explain for both analyzed RSGs. Figures~\ref{fig:hd-956872} and \ref{fig:v602-car2} show this new best fit with $\dot{M}=10^{-5.50}$ $M_{\odot}/\mathrm{yr}$ for HD~95687 and $\dot{M}=10^{-5.0}$ $M_{\odot}/\mathrm{yr}$ for V602~Car and the newly found parameters of the $\beta$-law $\beta_{\mathrm{new}}=-1.60$ and $\gamma_{\mathrm{new}}=0.05$ (green) in contrast with the initial best fit (red). We should point out that the best fitted $\beta$-law parameters $\beta_{\mathrm{new}}=-1.60$ and $\gamma_{\mathrm{new}}=0.05$ are the limit of the grid, however, if we go to higher values the $\beta$-law stops having smooth shape, so we would have troubles computing the radiative transfer equations with {\sc Turbospectrum}.

%----------------------------------------------------------------------

The velocity profile is found assuming a fiducial wind limit of $v_{\infty}=25\pm5$ km/s, that is the value matched to \citet{1998IrAJ...25....7R,2005A&A...438..273V,10.1093/mnras/stx3174}, and Equation~\ref{eq:masscont}. We assume no velocity gradient since the acceleration region is shallow and $v_{\infty}$ is due to turbulent motions \citep[see][]{2021MNRAS.508.5757D}. The model is sensitive to the density $\rho$, meaning that $\dot{M}$ and $v$ are degenerate with one another. 
 %We study the effect of changing the $v_{\infty}$ in Section~\ref{sub:vinf}.

%--------------------------------------------------------------------------

For the temperature profile we first used simple radiative transfer equilibrium (R.E.), defined as:
\begin{equation}\label{eq:re}
T(r_{\mathrm{wind}})=T(\tau_{\mathrm{Ross}}=2/3)\sqrt{R_{\star}r_{\mathrm{wind}}}
\end{equation}
where T($\tau_{\mathrm{Ross}}$=2/3) and $R_{\star}$ are the temperature and radius at the bottom of the photosphere, and $T(r_{\mathrm{wind}})$ and $r_{\mathrm{wind}}$ are the temperature and radius of the wind extension, respectively. This will result in a smoothly decreasing temperature profile for the extended atmosphere.

%- Evidence for chromospheric temperature inversion 
In addition, \citet{2021MNRAS.508.5757D} defined a different temperature profile in their semi-empirical 1D model of $\alpha$~Ori by \citet{2001ApJ...551.1073H}, based on spatially-resolved radio continuum data. The main characteristic of this profile is a temperature inversion in the chromosphere of the star, that peaks at $\sim 1.4\,R_{\star}$, and decreases again. ALMA and VLA observations of the RSGs Antares and Betelgeuse by \citet{1998Natur.392..575L,2017A&A...602L..10O,2020A&A...638A..65O} confirm the presence of such a lukewarm chromospheric temperature inversion, peaking at a radius of $1.3-1.5\,R_{\star}$ with a peak temperature of $\sim$3800 K. However, \citet{1998Natur.392..575L} pointed out that optical and ultraviolet chromospheric signatures required higher temperatures of $\sim$5000 K at similar radii \citep{1996BAAS...28..942U}. Conversely, modelling of spectroscopic and interferometric data of the CO MOLsphere derived gas temperatures of only $~$2000 K at $1.2-1.4\,R_{\star}$ \citep{2013A&A...555A..24O}. \citet{2020A&A...638A..65O} suggested these components co-exist in different structures at similar radii in an inhomogeneous atmosphere, and are spatially unresolved by current measurements. Observations at different wavelengths may then be sensitive to different such structures.

Following \citet{2021MNRAS.508.5757D}, we include in our model setup either a temperature profile in R.E., which may be more relevant for observations of the near-IR MOLsphere, or a temperature profile with a chromospheric temperature inversion which may be more relevant for chromospheric  signatures in the optical or UV or for radio continuum observations.

\medskip
%----------------------------------------------------------------------------
Once the density, temperature and velocity profiles are defined, we re-sample the model to a constant logarithmic optical depth sampling $\Delta \log(\tau)$, and we use $0.01<\Delta \log(\tau)<0.05$. The reasons for this re-sampling are explained in \citet{2021MNRAS.508.5757D}: if the grid is too finely sampled, rounding errors can occur leading to numerical difficulties. On the other hand, if the sampling is too coarse, the $\tau_{\lambda}=2/3$ surface is poorly resolved for strong absorption lines. %In Section~\ref{sub:tau} we explore the limitations of this re-sampling to constant $\Delta \log \tau$.

Finally, we define the outer boundary of the model where the local temperature is $<800$ K, which is reached at $\sim8.5\,R_{\star}$. Below this temperature our code is unable to reliably converge the molecular equilibrium. In addition some species would be depleted to dust grains. %Below this temperatures there will be no stable molecular equilibrium, and certain molecular species will begin depleting to dust grains. %We must point out however, that recent studies suggest the formation of dust at earlier temperature layers than $\sim900$ K \citep{2020A&A...644A.139G}. \citet{2020A&A...644A.139G} study in the infrared the spectra of nine supergiants, and suggest the presence of silicate dust growth at a temperature of $920$ K. This means that our semi-empirical model would not be free from the effect of stellar dust, and this could cause a miss-match between observations and models. To explore this effect, we would have to model the effect of the formation of dust grains below $8.5\,R_{\star}$, which is out of the scope of this paper.

\begin{figure}[t]
\centering
\includegraphics[width=1.\linewidth]{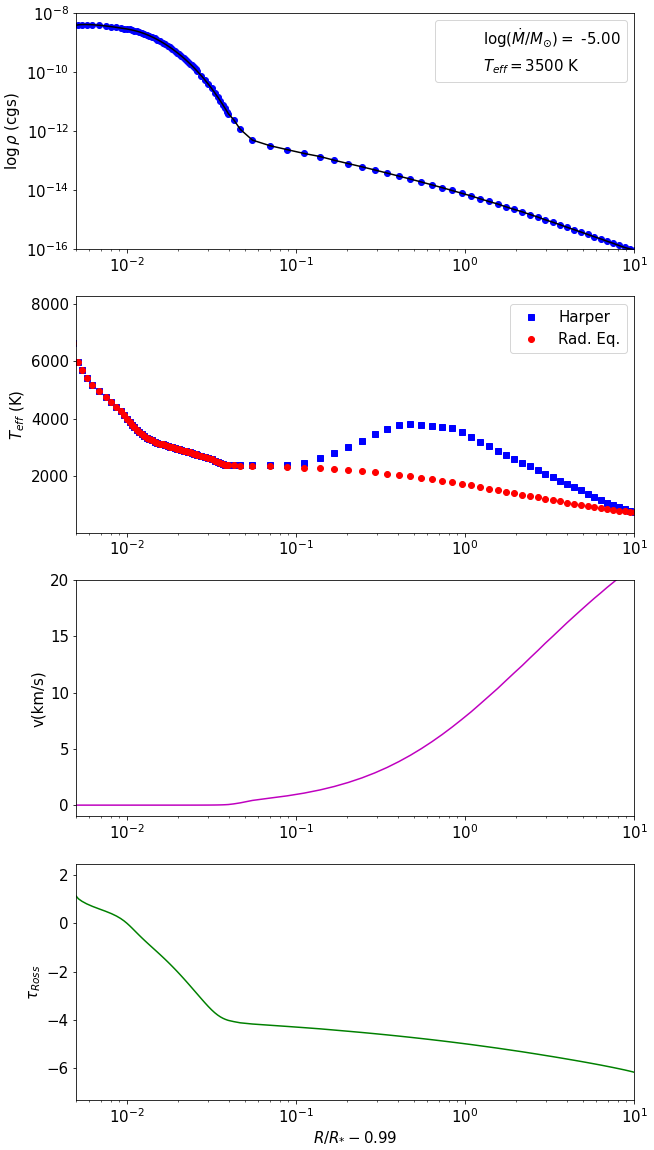}
\caption{\textit{From top to bottom:} The extended profiles for the density (blue dots), temperature (blue squares for temperature inversion by \citet{2001ApJ...551.1073H} a.k.a "Harper", and red circles for simple radiative equilibrium), wind velocity (purple) and Rosseland optical depth (green). The extended model with $\log \mathrm{L}/\mathrm{L}_{\odot}=4.8$,  $T_{\mathrm{eff}}=3500$ K, $\log g=0.0$, $M=15\,M_{\odot}$, $\dot{M}=10^{-5.5}$ $M_{\odot}/\mathrm{yr}$ and $R_{\mathrm{max}}=8.5\,R_{\star}$, corresponds to the stellar parameters of HD~95687.}
\label{fig:model}
\end{figure}

Figure~\ref{fig:model} shows the density, temperature, velocity and Rosseland opacity profiles for the example of an extended model with $\log \mathrm{L}/\mathrm{L}_{\odot}=4.8$,  $T_{\mathrm{eff}}=3500$ K, $\log g=0.0$, $M=15\,M_{\odot}$ and $\dot{M}=10^{-5.5}$ $M_{\odot}/\mathrm{yr}$. %The second panel shows the temperature profile with both the chromospheric temperature inversion (blue squares) and simple R.E. (red circles).   %To explore this effect, we have compared both methods in Section~\ref{sub:harp}, and found out that the visibility and SED profiles are better reproduced with R.E. instead.
%Figure 1: the model parameters

%\subsubsection{$M_{\odot}$ and mass-luminosity dependence}

\subsection{Computation of model intensities}

We computed both the spectra and the intensity profiles with respect to $\mu$, where $\mu=\cos \theta$, with $\theta$ being the angle between the radial direction and the emergent ray, and $\cos \theta=1$ corresponding to the intensity at the centre of the disk. For this we used the radiative transfer code {\sc turbospectrum v19.1} \citep{2012ascl.soft05004P}. Setting a wavelength range from $1.8\,\mathrm{\mu m}$ to $5.0\,\mathrm{\mu m}$  with a step of $0.1 \AA$, to explore the spectral range of the following instruments at the VLTI:
\begin{itemize}
\item GRAVITY \citep{2017A&A...602A..94G} for the $K$-band $1.8-2.5\,\mathrm{\mu m}$. 
\item MATISSE \citep{2022A&A...659A.192L} for the $L$ ($3.2<\lambda<3.9\,\mathrm{\mu m}$) and $M$-bands ($4.5<\lambda<5\,\mathrm{\mu m}$). We did not use the $N$-band ($8<\lambda<13\,\mathrm{\mu m}$) because it is dominated by dust emission. 
\end{itemize}%We use a uniform turbulent velocity of 2 km/s following \citet{2013A&A...554A..76A}, 

For the spectral synthesis, we included a list of atomic and molecular data. Chemical equilibrium is solved for 92 atoms and their first 2 ions, including Fe, Ca, Si and Ti, and molecular data for CO, TiO, H$_{2}$O, OH, CN  and SiO is included, among about 600 species.

The spectra and visibilities can be convolved to any spectral resolution used by GRAVITY and MATISSE, or instruments at other interferometers. In this work, we show as an example the results convolved to match the HIGH spectral resolution of MATISSE: $R=1000$\footnote{The full spectral resolution models will be available at CDS.}.%The results can be convolved to match any other spectral resolution of the VLTI instruments.

\subsection{Computation of model interferometric visibilities}
To compute the visibility from the intensity profile, we used the following Hankel transform as in \citet{2000MNRAS.318..387D}, 
\begin{equation}\label{eq:hankel}
V_{\mathrm{Model}}(\lambda)= \int_{0}^{1} S_{\lambda} I_{\lambda}^{\mu} J_{0}[\pi \theta_{\mathrm{Model}}(B/\lambda)(1-\mu^{2})^{1/2}]\mu d\mu
\end{equation}
where $V_{\mathrm{Model}}$ is the visibility of our model, $S_{\lambda}$ the instrument sensitivity curve, $I_{\lambda}^{\mu}$ the computed intensities with respect to $\mu$ from {\sc Turbospectrum}, $J_{0}$ is the 0th order Bessel function, $\theta_{\mathrm{Model}}$ is the angular diameter of the outermost layer of the model, and $B$ is the baseline of the observation. The $V_{\mathrm{Model}}$ is then normalised with respect to the total flux.

To estimate $\theta_{\mathrm{model}}$, we used the relation with the Rosseland angular diameter $\theta_{\mathrm{Ross}}$,
\begin{equation}\label{eq:theta}
\theta_{\mathrm{Ross}}=\frac{R(\tau_{\mathrm{Ross}}=2/3)}{R_{\mathrm{max}}}\theta_{\mathrm{Model}}
\end{equation}
found in \citet{2000MNRAS.318..387D} and \citet{2004A&A...413..711W}, where $R(\tau_{\mathrm{Ross}}=2/3)$ is the radius of the star at $\tau_{\mathrm{Ross}}=2/3$, defined as the photospheric layer, and  $R_{\mathrm{max}}$ is the outer most radius of our model.% $\theta_{\mathrm{Ross}}$ is the Rosseland-mean photospheric angular diameter. 

We use the definition in \citet{2017A&A...597A...9W} to scale the final visibility of the model as
\begin{equation}\label{eq:dust}
V(A,\theta_{\mathrm{Ross}})=A*V_{\mathrm{Model}}(\theta_{{\mathrm{Ross}}})
\end{equation}
where $A$ allows for the attribution of a fraction of the flux to an over-resolved circumstellar component \citep{2013A&A...554A..76A}, and $V_{\mathrm{Model}}(\theta_{\mathrm{Ross}})$ is the model visibility computed using Equation~\ref{eq:hankel} with an associated Rosseland angular diameter $\theta_{\mathrm{Ross}}$ from Equation~\ref{eq:theta}.

%--------------------------------------------------------------------

\section{Results}\label{sec:results}
%We first run {\sc Turbospectrum}, which allows for the computation of the SED and the intensities. 
\subsection{Base model}\label{sub:base}

We compute the spectra, intensities and $|V|^{2}$ for a base model of $T_{\mathrm{eff}}=3500$ K, $\log g=0.0$, $[Z]=0$, $\xi=5$ km/s, $M=15\,M_{\odot}$, $R_{\star}=690\,R_{\odot}$ and $R_{\mathrm{max}}=8.5\,R_{\star}$, corresponding to a RSG similar to HD~95687 \citep{2015A&A...575A..50A}. The density parameters in Equation~\ref{eq:betalaw} are $\beta_{\mathrm{Harp}}=-1.10$ and $\gamma_{\mathrm{Harp}}=0.45$ as in \citet{2001ApJ...551.1073H} and the wind limit $v_{\infty}=25$ km/s. The temperature profile is initially set to simple R.E., as we are interested in the near-IR $K$-band MOLsphere (cf. Section~\ref{sec:models}). However, we also look at the effect of a chromospheric temperature inversion in Section~\ref{sub:harp}. We use mass-loss rates of $\dot{M}=10^{-4}$, $10^{-5}$, $10^{-6}$ and $10^{-7}$ $M_{\odot}/\mathrm{yr}$, and a simple {\sc MARCS} model without any wind. As an example, we simulate a star with $\theta_{\mathrm{Ross}}=3$ mas, a baseline of $B=60$ m and without any additional over-resolved component, i.e. $A=1$. 

\begin{figure*}[t]
\centering
\includegraphics[width=1.\linewidth]{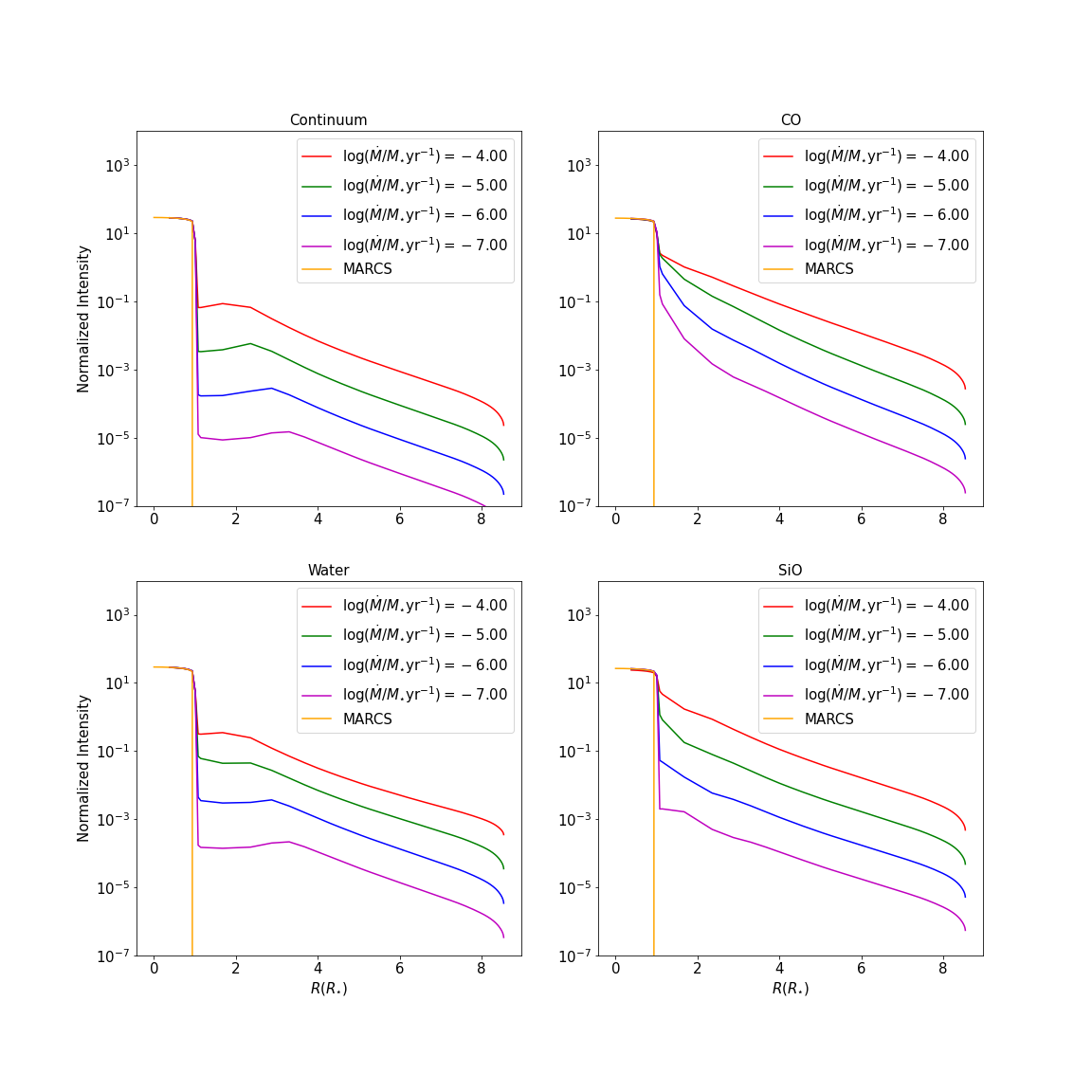}
\caption{Intensity with respect to the $R_{\mathrm{max}}$ of our model, for simple {\sc MARCS} (orange), $\dot{M}=10^{-4}$ (red), $10^{-5}$ (green), $10^{-6}$ (blue) and $10^{-7}$ $M_{\odot}/\mathrm{yr}$ (purple) and different wavelength cuts corresponding to: the continuum ($2.26\,\mathrm{\mu m}<\lambda<2.28\,\mathrm{\mu m}$, \textit{upper left}), the transition {\sc CO (2-0)} ($\lambda=2.29\,\mathrm{\mu m}$, \textit{upper right}), water ($1.9\,\mathrm{\mu m}<\lambda<2.1\,\mathrm{\mu m}$, \textit{lower left}) and SiO ($\lambda=4.0\,\mathrm{\mu m}$, \textit{lower right}). We observe an extension in all cases except for the {\sc MARCS} model without the addition of a wind. }
\label{fig:i1d}
\end{figure*}

Figure~\ref{fig:i1d} shows the intensities with respect to the extended stellar radius $R_{\mathrm{max}}$ for a cut in the continuum ($2.26\,\mathrm{\mu m}<\lambda<2.28\,\mathrm{\mu m}$), the transition {\sc CO (2-0)} ($\lambda=2.29\,\mathrm{\mu m}$), water ($1.9\,\mathrm{\mu m}<\lambda<2.1\,\mathrm{\mu m}$) and SiO ($\lambda=4.0\,\mathrm{\mu m}$) for the different $\dot{M}$. %for a simple MARCS model, $\dot{M}=10^{-4}$, $10^{-5}$, $10^{-6}$ and $10^{-7}$ $M_{\odot}/\mathrm{yr}$. 

%We have also computed the 2D images of our stellar models for MARCS, $\dot{M}=10^{-4}$, $10^{-5}$, $10^{-6}$ and $10^{-7}$ $M_{\odot}/\mathrm{yr}$ for the case of {\sc CO (2-0)}, which is one of the most extended molecules as seen in Figure~\ref{fig:i1d}. %The 2D images are shown in Figures~\ref{fig:2dm}, \ref{fig:2d7}, \ref{fig:2d6}, \ref{fig:2d5} and \ref{fig:2d4}. 

We observe an extension in all cases except for the {\sc MARCS} model without the addition of a wind. The CO lines  and the SiO lines seem to have the most prominent presence throughout the extended atmosphere (highest intensity compared to water or the continuum). %The same can be seen in Figures~\ref{fig:2dm}, \ref{fig:2d7}, \ref{fig:2d6}, \ref{fig:2d5} and \ref{fig:2d4}: compared to simple MARCS (Figure~\ref{fig:2dm}) where there is only a slight limb darkening at $\sim1R_{\star}$, the rest of the Figures~\ref{fig:2dm}, \ref{fig:2d7}, \ref{fig:2d6}, \ref{fig:2d5} and \ref{fig:2d4} show extension up to $8.5R_{\star}$. 

\begin{figure*}[t]
\centering
\includegraphics[width=1.\linewidth]{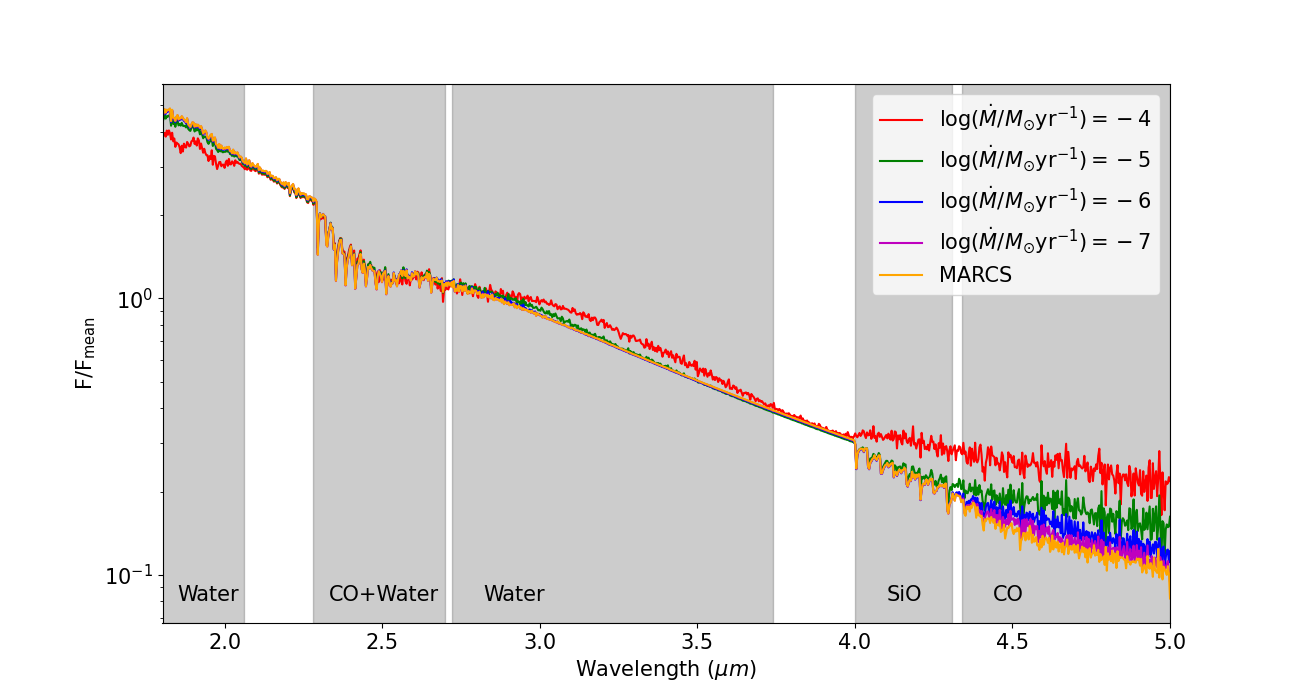}
\caption{Normalised model spectra with respect to the mean flux for $\dot{M}=10^{-4}$ (red), $10^{-5}$ (green), $10^{-6}$ (blue) and $10^{-7}$ $M_{\odot}/\mathrm{yr}$ (purple). We also plot the spectra and visibilities based on the {\sc MARCS} model without a wind (orange).This is the case for R.E. We have convolved the results with the spectral resolution of $R=1000$. The main differences within models can be seen in the water, CO and SiO molecular bands, the corresponding wavelength regions are highlighted in light grey.}
\label{fig:sed}
\end{figure*}

\begin{figure*}[t]
\centering
\includegraphics[width=1.\linewidth]{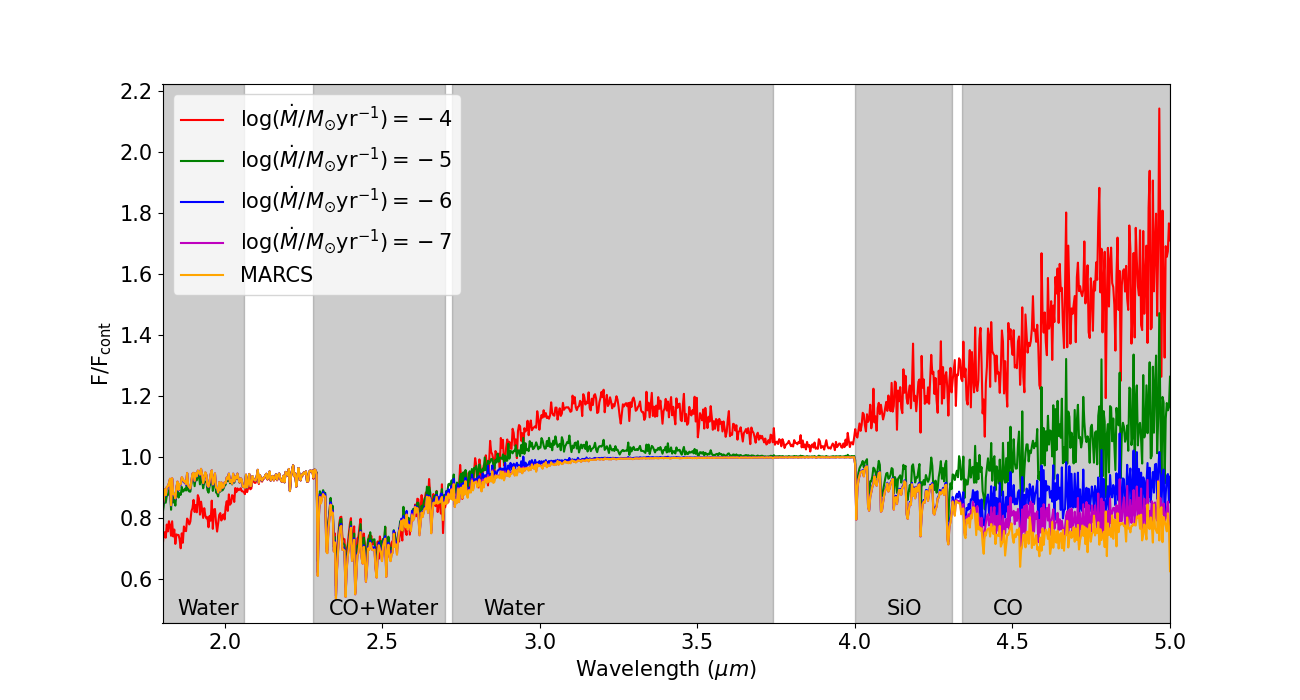}
\caption{Same as Figure~\ref{fig:sed} but for the flux normalised to the continuum.}
\label{fig:sedc}
\end{figure*}

\begin{figure*}[t]
\centering
\includegraphics[width=1.\linewidth]{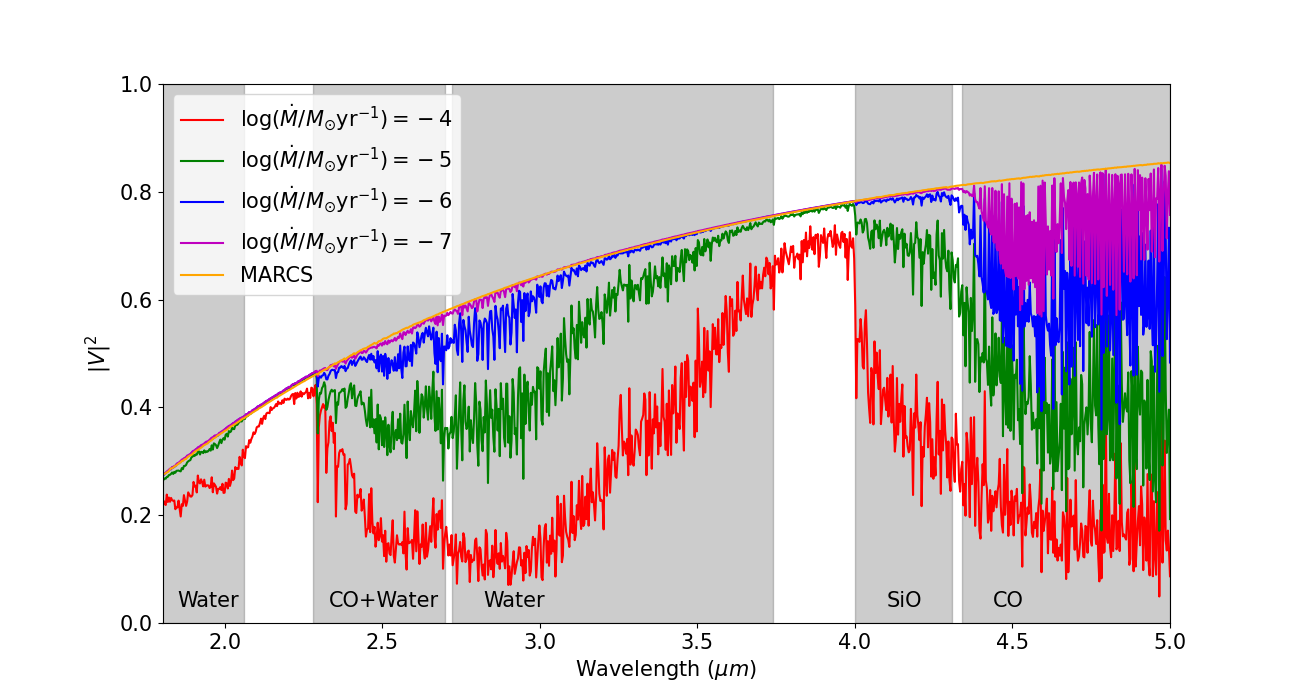}
\caption{Same as Figure~\ref{fig:sed} but for the modelled visibility $|V|^{2}$. The baseline assumed is $B=60$ m.}
\label{fig:v2}
\end{figure*}

Figures~\ref{fig:sed},\ref{fig:sedc} and \ref{fig:v2} show the spectra, the normalised spectra to the continuum, and squared visibility amplitudes ($|V|^{2}$) computed from our base model with the different $\dot{M}$, from highest $\dot{M}=10^{-4}$ $M_{\odot}/\mathrm{yr}$, to the simple {\sc MARCS} model without wind. 

%convolved with the spectral resolution of $R=1000$ are shown for different  $\dot{M}=10^{-4}$, $10^{-5}$, $10^{-6}$ and $10^{-7}$ $M_{\odot}/\mathrm{yr}$. We also plot the spectra and visibilities based on the {\sc MARCS} model without a wind. %The temperature profile in Figures~\ref{fig:sed} and \ref{fig:v2} is based on the simple R.E. since it can best reproduce the features of the case study as commented in Section~\ref{sec:analysis}. Initially, our density profile is computed with $\beta_{\mathrm{Harp}}=-1.10$ and $\gamma_{\mathrm{Harp}}=0.45$.

% For comparison, we also computed the SED and $|V|^{2}$ for a simple R.E (Figures~\ref{fig:harp} and \ref{fig:harp2}). We have found that there is little difference with both models unless we go to high mass-loss rates (Figure~\ref{fig:harp2}) where CO in emission appears for \citet{2001ApJ...551.1073H} but not for R.E..

%We have convolved the results with the resolution for each VLTI instrument, using the following resolutions as examples:
%\begin{itemize}
%\item PIONIER with $R\sim40$.
%\item GRAVITY with $R\sim4000$.
%\item MATISSE with $R\sim959$ (high) for the L-band, and $R\sim506$ (medium) for the M-band. we do not include the N-band because for very high wavelengths, the model will fail since we do not include the effect of dust. 
%\end{itemize}

%Figure 2: model with R~1000 for SED and V2

When comparing the spectra and $|V|^{2}$ in the Figures~\ref{fig:sed}, \ref{fig:sedc} and \ref{fig:v2}, there are several things to notice:
\begin{itemize}
 \item In Figures~\ref{fig:sed} and \ref{fig:sedc}, the spectral signatures of CO in the wavelength range of $\lambda=2.29-2.7\,\mathrm{\mu m}$ ($K$-band) do not strongly depend on the $\dot{M}$, as they remain relatively unchanged. Only at high $\dot{M}$ and high resolution spectra the low excitation lines start becoming stronger, as predicted by \citet{1988A&A...197..185T}. %This means that for higher mass-loss rates the CO will stop being absorved and will start to emit.
\item In the region $\lambda>4.0\,\mathrm{\mu m}$ ($L$ and $M$-bands), the spectra shows the presence of SiO lines at wavelengths up to $\sim4.3\,\mathrm{\mu m}$, which seem to remain in absorption up to high $\dot{M}$ ($\sim10^{-4}$ $M_{\odot}/\mathrm{yr}$). At $\gtrsim4.3\,\mathrm{\mu m}$ we observe the presence of CO as we increase the $\dot{M}$. The CO lines in the $M$-band are observed in emission already at low $\dot{M}$ (starting at $\dot{M}=10^{-6}$ $M_{\odot}/\mathrm{yr}$), while the CO at the $K$-band remains in absorption. 
\item In the $|V|^{2}$ (Figure~\ref{fig:v2}), the most important thing to notice is that the extended molecular layers, mostly of the CO lines in $\lambda=2.29-2.7\,\mathrm{\mu m}$ ($K$-band), are seen in the extended models. This extension was not reproduced earlier with {\sc MARCS} or {\sc PHOENIX} \citep{1999JCoAM.109...41H} models alone.
\item The CO extension increases with increased $\dot{M}$ (in all $K$, $L$ and $M$-bands). The atmospheric extension is best observed in the $M$-band, as we see a drop of $|V|^{2}$ already for a low mass-loss rate of $\dot{M}=10^{-7}$ $M_{\odot}/\mathrm{yr}$ (in purple) as compared to a simple {\sc MARCS} model with no extension (in orange). 
\item The visibility spectra are indicative of extended layers of water vapour (centred at $2.0\,\mathrm{\mu m}$) for high $\dot{M}$ ($\gtrsim \dot{M}=10^{-5}$ $M_{\odot}/\mathrm{yr}$), while the flux spectra are less sensitive. This water features are present in the observations of RSGs \citep[e.g.,][]{2015A&A...575A..50A}.
\item Checking the $|V|^{2}$, we are not able to reproduce some atomic lines in the $2.10\,\mathrm{\mu m}<\lambda<2.30\,\mathrm{\mu m}$ region for mass-loss rates $\dot{M}<10^{-4}$ $M_{\odot}/\mathrm{yr}$, which are the most sensitive to the stratification very close to the stellar surface \citep{2020A&A...642A.235K}. %The description very close to the stellar surface may not be optimum yet. %for very high mass losses ($\gtrsim \dot{M}=10^{-5}$ $M_{\odot}/\mathrm{yr}$) there is the presence of water in the wavelength range $\lambda=1.8-2.2\,\mathrm{\mu m}$. This water feature is smaller in the spectra and only present in high $\dot{M}$. 
 %However, we observe the water presence in the $|V|^{2}$ of our VLTI/AMBER case studies (lower left panels in Figures~\ref{fig:hd-95687} and \ref{fig:v602-car} in grey). This means that either our best fitted mass-loss rate is $\gtrsim\dot{M}=10^{-5}$ $M_{\odot}/\mathrm{yr}$ which would not be in accordance with other mass-loss rate prescriptions \citep{1988A&AS...72..259D,2005ApJ...630L..73S,2020MNRAS.492.5994B}, or our model fails to reproduce this water feature for more reasonable mass-loss rates. 
\end{itemize}

\subsection{Variations of the base model}\label{sub:var}
%--------------------------------------------------------------
\subsubsection{Density profile}\label{sub:beta}
%The fact that the temperature profile is not the cause of the missing water feature in $|V|^{2}$ for low mass-loss rates is not surprising, as we already have mentioned in Section~\ref{sec:analysis} that we assume the same extended T(r) for all $\dot{M}$. If we are seeing the water feature for high $\dot{M}$, while the temperature profile remains the same, this means that the most notable difference should occur in the density profile instead. 

\begin{figure}[t]
\centering
\includegraphics[width=1.\linewidth]{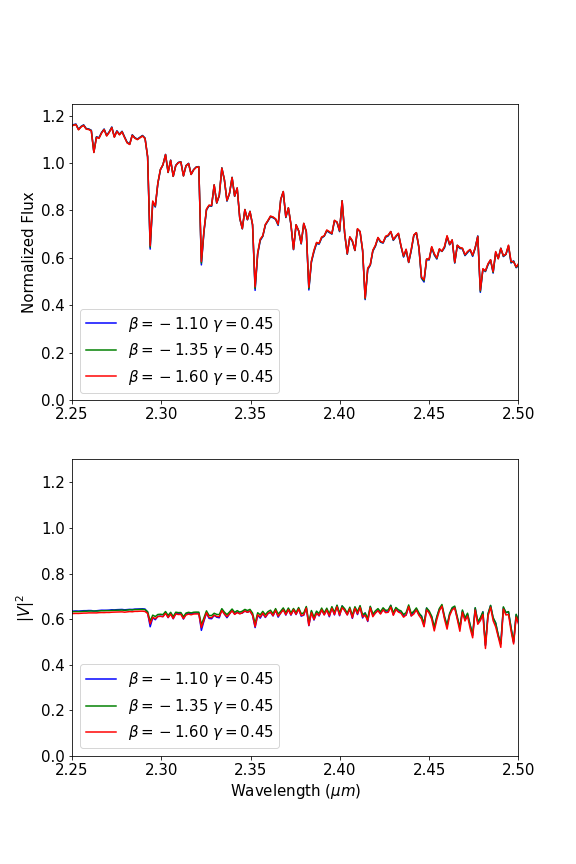}
\caption{Normalised model spectra (\textit{upper panel}) and $|V|^{2}$ (\textit{lower panel}) for a fixed $\gamma=0.45$ and the different $\beta=-1.10$ (blue), $-1.35$ (green) and $-1.60$ (red) in the K-band. The baseline used is $B=63.8$ m, corresponding to the case study of HD~95687 in Section~\ref{sec:case}. We can see that as the $\beta$ gets lower, the water features in $\lambda=2.35-2.5\,\mathrm{\mu m}$ become slightly more prominent.}
\label{fig:kband}
\end{figure}

So far we have assumed the density parameters (Equation~\ref{eq:betalaw}) from \citet{2001ApJ...551.1073H}: $\beta_{\mathrm{Harp}}=-1.10$ and $\gamma_{\mathrm{Harp}}=0.45$.  %However, the CO is formed at the outer region of the extended atmosphere, in contrast the water is formed in closer layers to the stellar photosphere \textbf{reference?}. %As seen in Figure~\ref{fig:beta},the model with $\log \dot{M}/M_{\odot}=-4.0$  (blue) shows in the $|V|^{2}$ a clear  presence of water at $\lambda=1.8-2.2\,\mathrm{\mu m}$ (upper panel in Figure~\ref{fig:v2}). The case of $\log \dot{M}/M_{\odot}=-5.5$  (red in Figure~\ref{fig:beta}), the water presence cannot be found since it stops being distinguishible at $\lesssim\log \dot{M}/M_{\odot}=-5.00$ in the $|V|^{2}$ (second panel n Figure~\ref{fig:v2}).
%Checking $|V|^{2}$ we see that the CO seems to be less extended for lower $\dot{M}$, whereas water is not present unless we go to unreasonably high $\dot{M}$. A possible explanation for this CO and water behaviour could be the following: the density closer to the photosphere should be as predicted for higher mass-loss rates, while the density at the end of the extension remains the same as predicted for lower mass-loss rates. In short, the density profile is shallower than what \citet{2001ApJ...551.1073H} shows. %We need to study the effect on $\beta$ and $\gamma$ in the density profile. 
%\textbf{REDO} To explore the variations of the density profile close to the stellar surface, we have used a grid of $-1.1<\beta<-1.60$ in steps of $\Delta\beta=0.25$ and $0.05<\gamma<0.45$ in steps of $\Delta\gamma=0.2$ (Figure~\ref{fig:beta-law-grid}). 
%Checking the K-band, ADD PLOTS OF SED AND V2 AND DISCUSS. WATER
Figure~\ref{fig:kband} shows the spectra and $|V|^{2}$ for the $K$-band for the different $\beta$ values defined in Figure~\ref{fig:beta-law-grid}. We did not include the variations on $\gamma$ because these produce an almost identical plot. Although the spectra remain unchanged by variations of $\beta$, the $|V|^{2}$ change slightly: as the density profile gets steeper (i.e. lower $\beta$ or $\gamma$ values), the extension features due to water in $\lambda=2.35-2.5\,\mathrm{\mu m}$ become more prominent. This can be understood as water layers form close to the stellar surface \citep{2020A&A...642A.235K}, and therefore are sensitive to variations of the density profile in this region. The measurements by \citet{2001ApJ...551.1073H} and their constraints of $\beta$ and $\gamma$ were less sensitive to the region very close to the stellar surface. %As mentioned, this is because the shallow density profile gives higher values closer to the photosphere where the water is formed, so a profile similar to initial high $\dot{M}$, as seen in Figure~\ref{fig:beta}. %The CO extension at $\lambda=2.3-2.5\,\mathrm{\mu m}$ also seems to increase with shallower profiles, which would explain  

\begin{figure}[t]
\centering
\includegraphics[width=1.\linewidth]{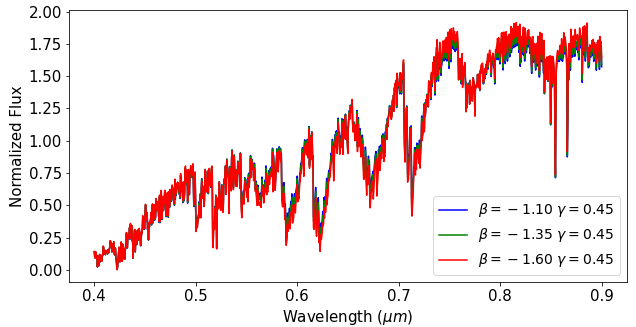}
\caption{Normalised model spectra for for a fixed $\gamma=0.45$ and the different $\beta=-1.10$ (blue), $-1.35$ (green) and $-1.60$ (red) in the optical TiO band region. As we decrease $\beta$, the TiO bands deepen slightly. }
\label{fig:optical}
\end{figure}

Furthermore, in Figure~\ref{fig:optical} we show the spectra in the optical TiO region ($\lambda=0.5-0.75\,\mathrm{\mu m}$) to see the effect of changing the $\beta$ parameter in the TiO lines. Again, changing $\gamma$ parameter produces a very similar plot. A discussion concerning the general effect of this semi-empirical model on the TiO bands can be found in \citet{2021MNRAS.508.5757D}. Briefly, an increase of the mass-loss rate will cause the TiO absorption lines to deepen, shifting the star to later spectral types (e.g., for a zero-wind model of spectral type M0, if we apply $\dot{M}=10^{-6}$, $\dot{M}=10^{-5.5}$ and $\dot{M}=10^{-5}$ $M_{\odot}/\mathrm{yr}$ in our model the star will be classified as M1, M2 and >M5, respectively). Changing the density profile parameters with a fixed $\dot{M}$ will affect the TiO bands, also shifting the stellar classification slightly to later spectral types as we deepen the TiO bands (Figure~\ref{fig:optical}). 

On the other hand, the TiO bands may be more sensitive to a higher chromospheric temperature component than the molecular layers in the near-IR, which may cause the TiO lines to be less deep (cf. Section~\ref{sec:models}).%In our case, the best parameters to fit the water shape seem to increase the TiO bands in the optical, causing the star classification to move to later spectral types. 

%-----------------------------------------------
\subsubsection{Temperature profile}\label{sub:harp}

\begin{figure}[t]
\centering
\includegraphics[width=1.\linewidth]{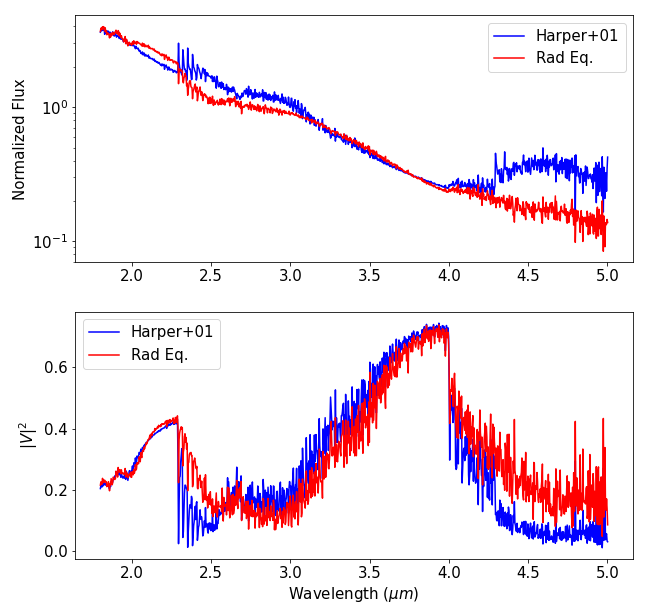}
\caption{Normalised model spectra (\textit{upper panel}) and $|V|^{2}$ (\textit{lower panel}) for our base model with simple radiative equilibrium (red) and the chromospheric temperature inversion profile by \citet{2001ApJ...551.1073H} (blue). Both models with a $\dot{M}=10^{-4.0}$ $M_{\odot}/\mathrm{yr}$. We can see that the main differences are in the regions $\lambda=2.35-3.0\,\mathrm{\mu m}$ and $\lambda>4.0\,\mathrm{\mu m}$ due to CO, where for the \textit{Harper+01} case the molecular lines appear more strongly in emission, and the visibilities are decreased due to a larger apparent diameter of the star.
 }
\label{fig:harp}
\end{figure}

\begin{figure}[t]
\centering
\includegraphics[width=1.\linewidth]{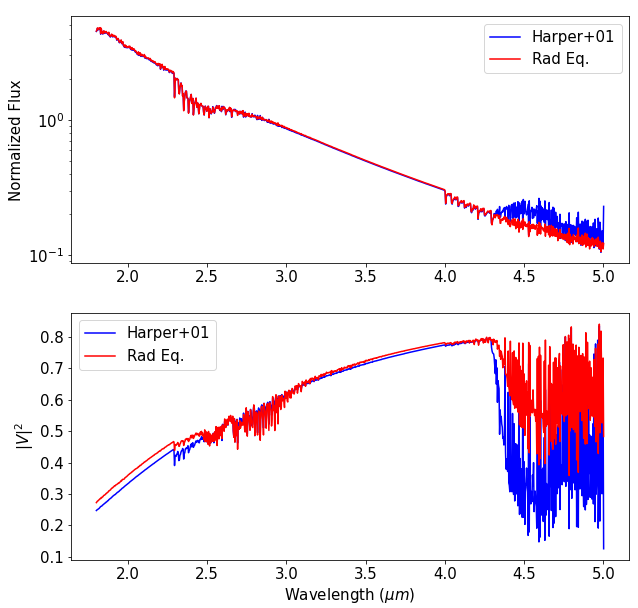}
\caption{Same as Figure~\ref{fig:harp}, but with a $\dot{M}=10^{-6.0}$ $M_{\odot}/\mathrm{yr}$. In this case, the main differences are only in the $\lambda>4.0\,\mathrm{\mu m}$ region due to CO, where for the \textit{Harper+01} case the molecular lines appear more strongly in emission, and the visibilities are decreased due to a larger apparent diameter of the star. }
\label{fig:harp2}
\end{figure}

%As mentioned, the previous results in Figures~\ref{fig:i1d}, \ref{fig:sed} and \ref{fig:v2} are based on R.E.  but we also have done the analysis with the temperature inversion profile by \citet{2001ApJ...551.1073H}. 
Figures~\ref{fig:harp} and \ref{fig:harp2} show the spectra and $|V|^{2}$ for our two temperature stratifications defined in Section~\ref{sec:analysis}, R.E and temperature inversion, with $\dot{M}=10^{-4}$ $M_{\odot}/\mathrm{yr}$ and $\dot{M}=10^{-6}$ $M_{\odot}/\mathrm{yr}$, respectively. 

When comparing both temperature profiles in Figures~\ref{fig:harp} and \ref{fig:harp2}, we see the main difference in the CO lines: in the $K$-band region  $\lambda=2.29-2.7\,\mathrm{\mu m}$ the CO is in emission when using the temperature inversion profile \citep[as also predicted by][]{2020A&A...638A..65O}, while for R.E. it remains in absorption even for very high $\dot{M}$ such as $\dot{M}=10^{-4}$ $M_{\odot}/\mathrm{yr}$. This difference can be seen in the lower panels of Figures~\ref{fig:harp} and \ref{fig:harp2}, where the R.E. shows less extension in the CO region as compared with the temperature inversion. For lower mass-loss rates ($\dot{M}=10^{-6}$ $M_{\odot}/\mathrm{yr}$, Figure~\ref{fig:harp2}), it gets harder to see the difference in both profiles. The only region that seems to make a difference is the $M$-band ($4.5<\lambda<5\,\mathrm{\mu m}$), where we see emission in the temperature inversion case. %The implications of this will be discussed in Section~\ref{sub:harp}.

Observations of CO lines in the $K$-band generally show CO in absorption, even for an extreme case such as the RSG VY~CMa \citep{2012A&A...540L..12W}, confirming that the lower temperature based on R.E. is better suited to describe the CO MOLsphere than the higher temperature components of the chromospheric temperature inversion (cf. Section~\ref{sec:models}).

%The cause of these peculiarities between models can be explained as follows: the chromospheric temperature inversion used by \citet{2001ApJ...551.1073H} will increase the $T_{\mathrm{eff}}$ in the chromospheric region. The first consequence is that some molecular lines will be depleted. Moreover, this will cause the CO lines to be re-emitted, while the R.E. will reproduce the absorption lines instead. In Section~\ref{sec:discussion} we present a further discussion on the temperature inversion results, and additional justifications why we end up using R.E. for our model and case study comparisons.

%Moreover, both R.E. and temperature inversion show the presence of water at $\lambda=1.8-2.2\,\mathrm{\mu m}$ only for high mass-loss rates $\gtrsim\log \dot{M}/M_{\odot}=-5.00$. This is the main problem when fitting our $|V|^{2}$ case study data to our model (see Section~\ref{sec:case}). The AMBER data has the water presence in the $|V|^{2}$, but to fit that we would have to go to unreasonably high mass-loss rates $\gtrsim\dot{M}=10^{-5}$ $M_{\odot}/\mathrm{yr}$ which would not be in accordance with the usual mass-loss rate prescriptions \citep[e.g.,][]{1988A&AS...72..259D,2005ApJ...630L..73S,2020MNRAS.492.5994B}, regardless of the temperature profile. We see that changing the temperature profile does not explain the fact that the water presence for modest mass-loss rates is not present. 

%Figure 4: Harper SED and V2

\bigskip
To sum up this Section~\ref{sec:results}, the {\sc MARCS}+wind model shows significant atmospheric extension in all wavelengths compared to a simple {\sc MARCS} model (Figure~\ref{fig:i1d}). Such an extension has  been observed, but it has so far not been reproduced by current models \citep{2013A&A...554A..76A}. %By extension meaning that the continuum and all other elements shown in Figure~\ref{fig:i1d} are present beyond the $\sim1R_{\star}$ photospheric radius, so the atmosphere shows extension in the studied wavelength ranges.
As we increase the $\dot{M}$, for a R.E. temperature profile the CO, SiO and water remain relatively unchanged in the spectra (Figure~\ref{fig:sed}), while for the $|V|^{2}$ we see a larger extension in all cases (Figure~\ref{fig:v2}). The R.E. seems to better reproduce the spectra than the temperature inversion as we do not observe the CO in emission in our case studies (see Section~\ref{sec:case}) nor other previously published data \citep[e.g.,][]{2012A&A...540L..12W,2013A&A...554A..76A,2015A&A...575A..50A}. Changing the $\beta$ and $\gamma$ parameters in Equation~\ref{eq:betalaw} deepens the water features in the $|V|^{2}$.

\section{Case study: comparison with HD~95687 and V602~Car}\label{sec:case}
In this section, we compare our model to published VLTI/AMBER data of the two RSGs HD~95687 and V602~Car available by \citet{2015A&A...575A..50A}. %\textbf{We have chosen these two targets because they are examples of two stars with different masses and henceforth different mass-loss rates, as well as two distinct luminosities.}
As previously mentioned, we have chosen these two RSGs since they sample different luminosities, $\dot{M}$ and masses. In addition, the data is readily available, and these are two well studied RSGs whose fundamental parameters (e.g., $T_{\mathrm{eff}}$, $\log g$, $\theta_{\mathrm{Ross}}$, $\log \mathrm{L}/\mathrm{L}_{\odot}$) are well known. The data were taken using the AMBER medium-resolution mode ($R \sim 1500$) in the $K-2.1\,\mathrm{\mu m}$ and $K-2.3\,\mathrm{\mu m}$ bands.  %The data reduction is explained in detail at \citet{2015A&A...575A..50A}. 
%The baselines are in $K-2.1\mathrm{\mu m}$  $B=63.8$ m for HD~95687 and $B=60.8$ m for V602~Car, while for $K-2.3 \mathrm{\mu m}$ $B=61.1$ m for HD~95687 and $B=63.2$ m for V602~Car.

The parameters used for each initial {\sc MARCS} model are shown in Table~\ref{tab:marcs}, following \citet{2015A&A...575A..50A}. HD~95687 is characterised by a smaller luminosity, mass and radius than V602~Car. In addition, HD~95687 shows a weaker atmospheric extension than V602~Car.

\begin{table*}
\caption{Parameters of the {\sc MARCS} models used for the analysis of each RSG. From left to right: luminosity $\log \mathrm{L}/\mathrm{L}_{\odot}$, effective temperature $T_{\mathrm{eff}}$, surface gravity $\log g$, metallicity [Z], microturbulence $\xi$, mass as in \citet{2015A&A...575A..50A}. The last column shows the radius of the star in the photosphere in solar units (defined at $\tau_{\mathrm{Ross}}=2/3$).}
\label{tab:marcs}
\small
\centering
\begin{tabular}{c c c c c c c c}

      RSG & $\log \mathrm{L}/\mathrm{L}_{\odot}$ & $T_{\mathrm{eff}}$ (K) & $\log g$ & [Z] & $\xi$ (km/s) & $M/M_{\odot}$ & $R_{\star}/R_{\odot}$ \\
     \hline \hline
	HD~95687 & 4.8 & 3500 & 0.0 & 0 & 5 & 15 & 690  \\
	V602~Car & 5.1 & 3400 & -0.5 & 0 & 5 & 20 & 1015  \\
     \hline
     
\end{tabular}

\end{table*}

%We point out that the microturbulence $\xi$ found by \citet{2013A&A...554A..76A} was 2 km/s instead of 5 km/s. However, there was no MARCS model available for $15M_{\odot}$ with $\xi=2$ km/s, but we have computed the intesities and SED with {\sc Turbospectrum} with a $\xi=2$ km/s instead. Anyway, we have seen that the choise of microturbulence from a $2-5$ km/s range makes little to no effect in the results \citep{2021MNRAS.505.4422G}. 

%The visibility parameters in Equations~\ref{eq:theta} and \ref{eq:dust} in \citet{2015A&A...575A..50A} are $\theta_{\mathrm{Ross}}=5.08\pm0.75$ mas and $3.17\pm 0.50$ mas, and $A=0.99$ and $1.0$ for HD~95687 and V602~Car respectively estimated for their PHOENIX models \citep{1999JCoAM.109...41H}. Since we are using different models, our parameters may be different from \citet{2015A&A...575A..50A}. 

To estimate both $\theta_{\mathrm{Ross}}$ and $A$, we compute for each studied model the $|V|^{2}$ as a function of their spatial frequency $B/\lambda_{o}$ where $\lambda_{o}$ corresponds to the continuum region $2.23-2.27\,\mathrm{\mu m}$. We have compared the model and data $|V|^{2}$, and found the best fitting $\theta_{\mathrm{Ross}}$ and $A$ by means of a $\chi^{2}$ minimisation.  %since they find $\theta_{\mathrm{Ross}}=5.08\pm0.75$ mas and $3.26\pm0.5$ mas with both a $A=1.0$ for V602~Car and HD~95687, respectively. %An example of the best-fit model found can be seen in Figure~\ref{fig:baseline} for HD~95687 for a mass-loss rate $\log \dot{M}/M_{\odot}=-5.50$. Note that this fitting should be done for each model that is compared with the data. %The rest of the fit for each model used produces a similar plot. %\textbf{We need to describe which grids of Mdot and the other parameters we used to find the best-fit model.}  Table~\ref{tab:atheta} shows the parameters found for each model used.
%\begin{table*}
%\caption{Best fitting $\theta_{\mathrm{Ross}}$ and $A$ for each model used to their corresponding RSG data and mass-loss rate.}
%\label{tab:atheta}
%\small
%\centering
%\begin{tabular}{c c c c c c c c c}

%      RSG & \multicolumn{2}{c}{MARCS} & \multicolumn{2}{c}{$\log (\dot{M}/M_{\odot})=-5$} & \multicolumn{2}{c}{$\log (\dot{M}/M_{\odot})=-4$}& \multicolumn{2}{c}{$\log (\dot{M}/M_{\odot})=-3.5$} \\
%      & $\theta_{\mathrm{Ross}}$ & A & $\theta_{\mathrm{Ross}}$ & A & $\theta_{\mathrm{Ross}}$ & A & $\theta_{\mathrm{Ross}}$ & A  \\
%     \hline \hline
%	AH Sco & 6.00 & 0.83  & 5.32 & 0.82 & 5.30 & 0.82 & 5.30 & 0.84  \\
%	UY Sct & 5.5 & 0.92  & 5.00 & 0.94 & 5.00 & 0.94 & 5.00 & 0.95  \\
%	KW Sgr & 4.0 & 0.99  & 3.60 & 0.99 & 3.50 & 0.99 & 3.4 & 0.99  \\
%     \hline
     
%\end{tabular}

%\end{table*}

%Figure 7: Model fit for A and theta_Ross
%\begin{figure}[t]
%\centering
%\includegraphics[width=1.\linewidth]{figures/Best_theta_ross_5p0_HD-95687-RE.png}
%\caption{Example of the best fitted parameters $\theta_{\mathrm{Ross}}$ and $A$.}
%\label{fig:baseline}
%\end{figure}
%

\begin{figure*}[t]
\centering
\includegraphics[width=1.\linewidth]{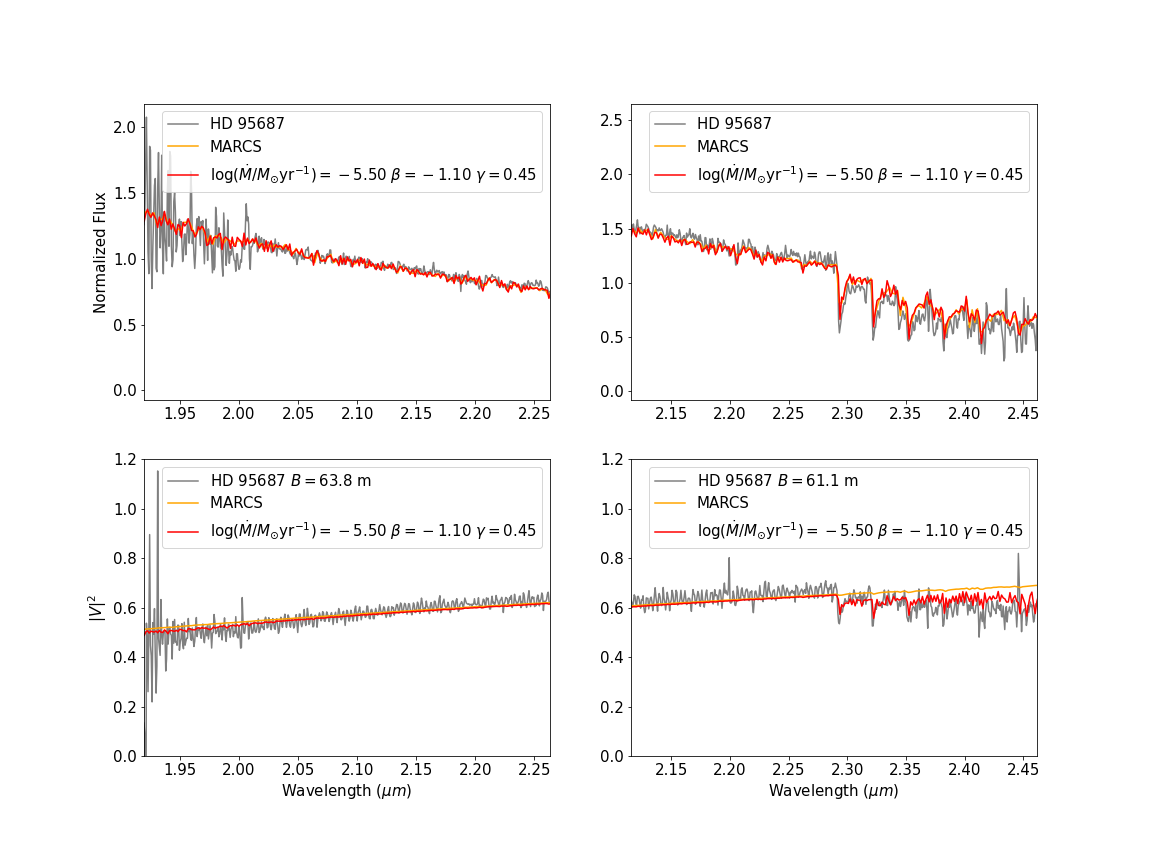}
\caption{\textit{Upper left: } Normalised flux for the RSG HD~95687 (grey), as observed with VLT/AMBER for the $K-2.1\,\mathrm{\mu m}$ bands. Our initial best-fit model with $\beta=-1.10$ and $\gamma=0.45$ is shown in red, corresponding to the parameters of the density profile by \citet{2001ApJ...551.1073H}. The pure {\sc MARCS} model fit is shown in orange. As expected, the fluxes are well represented by both our fit and {\sc MARCS}. \textit{Upper right: } Same as the upper left panel but for the $K-2.3\,\mathrm{\mu m}$ band. \textit{Lower left: } Same as the upper left panel but for the $|V|^{2}$ and a baseline of $B=60.8$ m. \textit{ Lower right:} Same as the lower right but for the $K-2.3\,\mathrm{\mu m}$ band and a baseline of $B=63.2$ m. Our model can represent better the data than simple {\sc MARCS}. }
\label{fig:hd-95687}
\end{figure*}

\begin{figure*}[t]
\centering
\includegraphics[width=1.\linewidth]{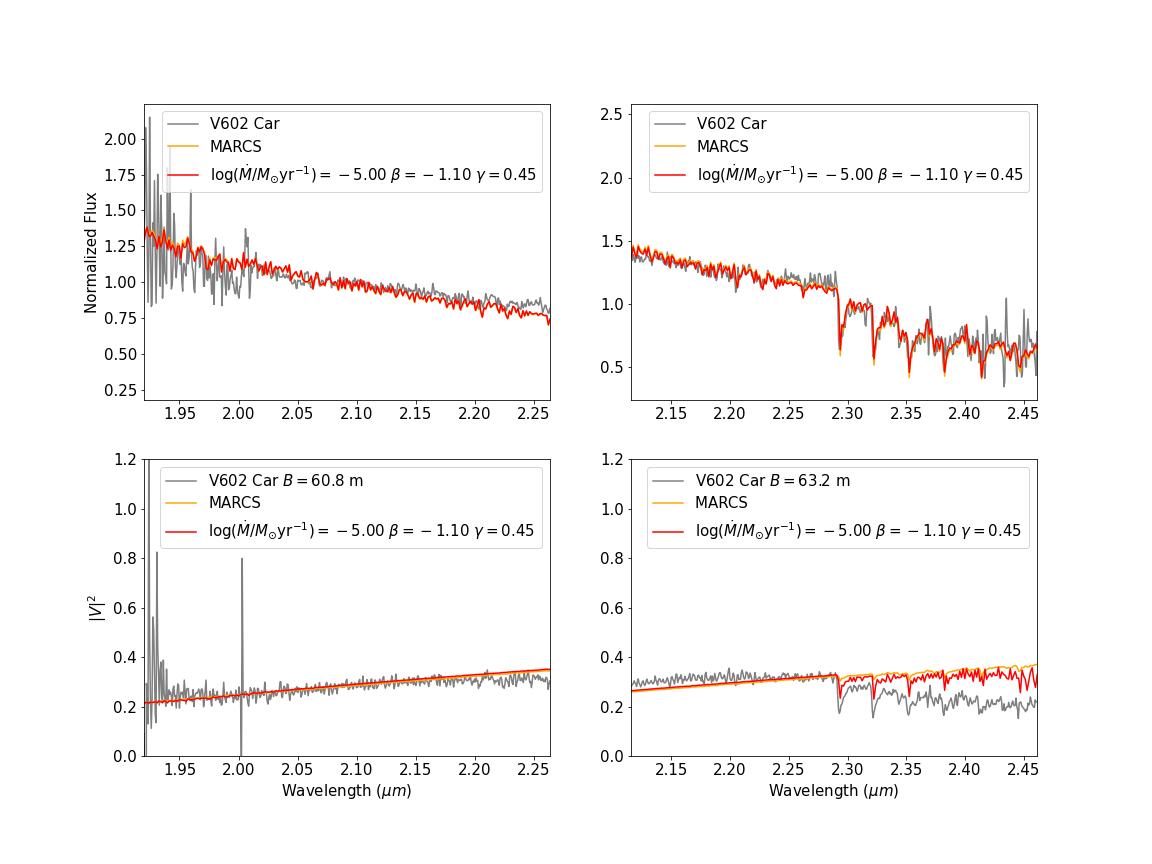}
\caption{Same as Figure~\ref{fig:hd-95687}, but for V602~Car.%Normalized flux and squared visibility amplitudes for the RSG V602~Car for only one baseline as observed with VLT/AMBER for the K1 and K2 bands in black, the best fitting results from our model and a simple MARCS model in orange.
}
\label{fig:v602-car}
\end{figure*}

Figures~\ref{fig:hd-95687} and \ref{fig:v602-car} show the {\sc MARCS} model fit to the data of \citet{2015A&A...575A..50A} and our initial {\sc MARCS}+wind model fit with $\beta_{\mathrm{Harp}}=-1.10$ and $\gamma_{\mathrm{Harp}}=0.45$, compared to the data of HD~95687 and V602~Car, respectively. For our model fit, we check both the spectra and $|V|^{2}$. We use a range of mass-loss rates of $-7<\log \dot{M}/M_{\odot}<-4$ with a grid spacing of $\Delta\dot{M}/M_{\odot}=0.25$. We obtain a best-fit of $\log \dot{M}/M_{\odot}=-5.50$ for HD~95687 and $\log \dot{M}/M_{\odot}=-5.0$ for V602~Car. These $\dot{M}$ are reasonable when compared with typical mass-loss prescriptions \citep[e.g.,][]{1988A&AS...72..259D,2005ApJ...630L..73S,2020MNRAS.492.5994B}.

%To find this value, we have done both a fit of the spectra and $|V|^{2}$ and a comparison to RSG mass-loss prescriptions \citep{1988A&AS...72..259D,2020MNRAS.492.5994B}.

For the temperature profile we use R.E., since the temperature inversion would either show depleted CO lines ($\lambda=2.29-2.7\,\mathrm{\mu m}$) for the spectra, which do not match with the observations, or not enough extension for the $|V|^{2}$. Therefore it is not possible to find a model with the temperature inversion profile that fits both spectra and $|V|^{2}$ simultaneously. %As mentioned in Section~\ref{sec:analysis}, this could be due to the presence of hot luke-warm chromospheres in co-spatial small cells of different temperatures at the same radii \citep{2020A&A...638A..65O}, whose effect would depend on the wavelength region we are studying. Therefore, the chromospheric temperatures would be too hot to affect $K$-band. %The implications of this are discussed in Section~\ref{sec:discussion}.

We estimate for our best fit models a $\theta_{\mathrm{Ross}}=5.35\pm0.7$ mas and $2.9\pm0.8$ mas, and $A=1.0\pm0.14$ and $1.0\pm0.08$ for V602~Car and HD~95687, respectively. The errors in $\theta_{\mathrm{Ross}}$ and $A$ are derived by the minimum values in the 68\% dispersion contours of the $\chi^{2}$ fit, that for 2 degrees of freedom corresponds to $\chi^{2}<\chi_{\mathrm{min}}^{2}+2.3$ \citep{1976ApJ...210..642A}. Our results are in agreement with \citet{2015A&A...575A..50A} within the error limits.
%It is important to notice in Figures~\ref{fig:hd-95687} and \ref{fig:v602-car} that our MARCS+wind model can match the data with better accuracy than a simple MARCS or PHOENIX models. Our model is therefore the only model that can reproduce both the spectra and $|V|^{2}$ extensions up to date. 

\medskip
In this work we show that, when adding a wind to a {\sc MARCS} model, we can now qualitatively fit the spectra and $|V|^{2}$. This is something that current existing models are unable to do. 

% taking into account that we cannot fit both SED and $|V|^{2}$ perfectly (see Section~\ref{sec:discussion}). We show in Figures~\ref{fig:hd-95687} and \ref{fig:v602-car} the best fit model for the case of simple radiative equilibrium; for both the SED (upper panels) and $|V|^{2}$ (lower panels), compared with the data (grey) for both K bands in VLTI/AMBER for HD~95687 and V602~Car, respectively. For clarity, we have only shown here only one baseline. %but the full $|V|^{2}$ fits for each baseline and each  extended models with mass losses $\log \dot{M}/M_{\odot}=-3.5$, $-4$, $-5$ and $-6$ are shown in Figures~\ref{fig:ah-sco2}, \ref{fig:uy-sct2} and \ref{fig:kw-sgr2} for AH-Sco, UY-Sct and KW-Sgr respectively.

\begin{figure*}[t]
\centering
\includegraphics[width=1.\linewidth]{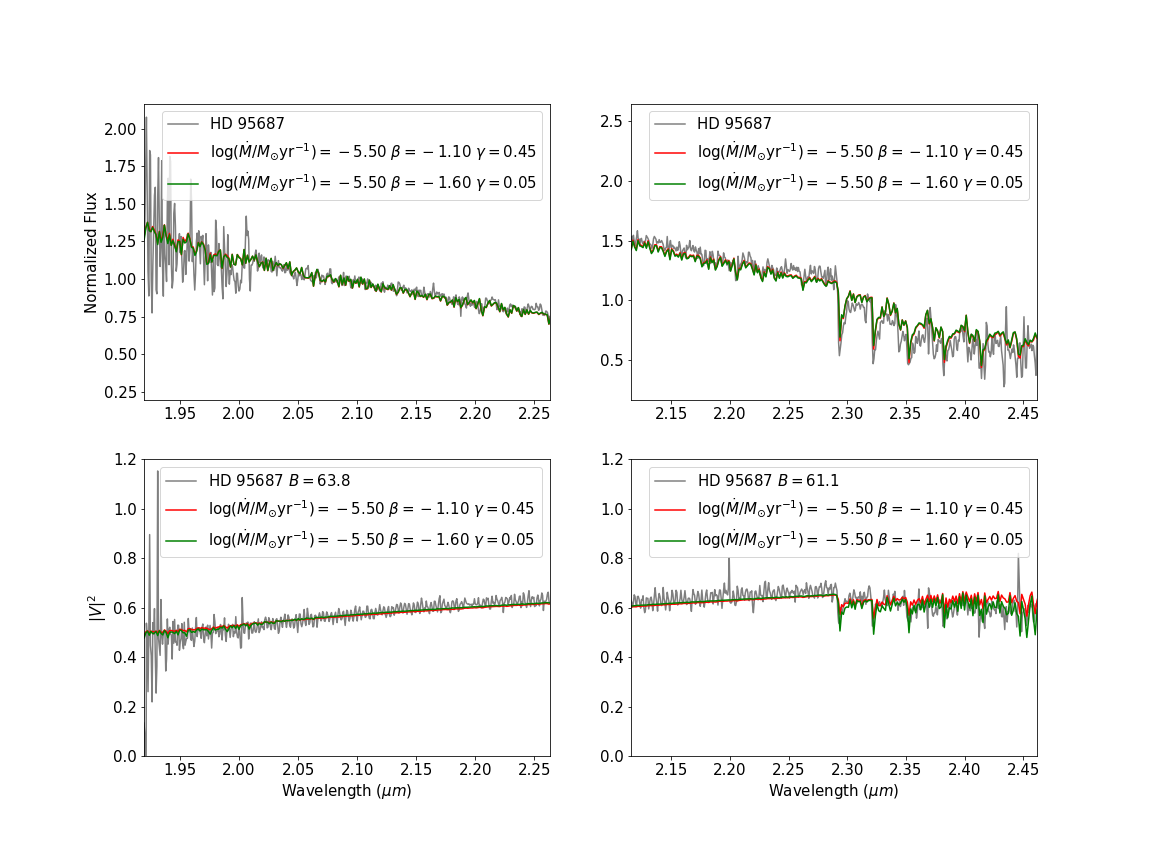}
\caption{ \textit{Upper left: } Normalised flux for the RSG HD~95687 (grey), as observed with VLT/AMBER for the $K-2.1\,\mathrm{\mu m}$ bands. Our initial best-fit model with $\beta=-1.10$ and $\gamma=0.45$ is shown in red, corresponding to the parameters of the density profile by \citet{2001ApJ...551.1073H}. The final best-fit model with $\beta=-1.60$ and $\gamma=0.05$ is shown in green. \textit{Upper right: } Same as the upper left panel but for the $K-2.3\,\mathrm{\mu m}$ band. \textit{Lower left: } Same as the upper left panel but for the $|V|^{2}$ and a baseline of $B=60.8$ m. \textit{Lower right:} Same as the lower right but for the $K-2.3\,\mathrm{\mu m}$ band and a baseline of $B=63.2$ m. The best fit model for $\beta$ and $\gamma$ can reproduce the water features in $\lambda=2.29-2.5\,\mathrm{\mu m}$ with better accuracy than our initial best fit.}
\label{fig:hd-95687-2}
\end{figure*}

\begin{figure*}[t]
\centering
\includegraphics[width=1.\linewidth]{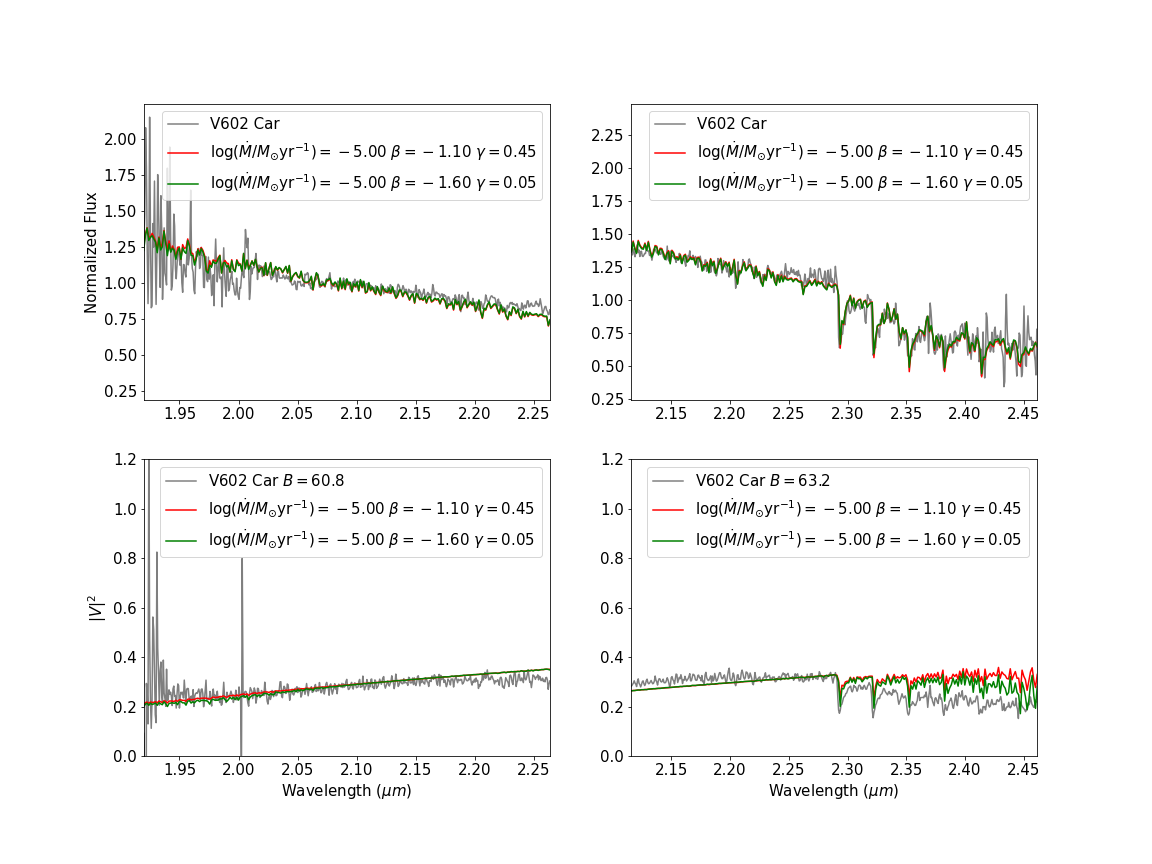}
\caption{Same as Figure~\ref{fig:hd-95687-2}, but for V602~Car.%Normalized flux and squared visibility amplitudes for the RSG V602~Car for only one baseline as observed with VLT/AMBER for the K1 and K2 bands in black, the best fitting results from our model and a simple MARCS model in orange.
}
\label{fig:v602-car-2}
\end{figure*}

\medskip
This initial fit can be further improved by modifying the inner wind density profile. %For this purpose, we use a grid with $-1.1<\beta<-1.60$ in steps of $\Delta\beta=0.25$ and $0.05<\gamma<0.45$ in steps of $\Delta\gamma=0.2$. We obtain a best-fit density parameters of $\beta=-1.60$ and $\gamma=0.05$ for both HD~95687 an V602~Car. 
Figures~\ref{fig:hd-95687-2} and \ref{fig:v602-car-2} show both the spectra and $|V|^{2}$ of the new fit changing the density parameters in comparison with the data and the initial {\sc MARCS}+wind fit for HD~95687 and V602~Car, respectively. The main difference between both {\sc MARCS}+wind models can be found in the $|V|^{2}$ water region at $\lambda=2.29-2.5\,\mathrm{\mu m}$, where the new $\gamma$ and $\beta$ fit better.

%We notice that the best fit $\beta$ and $\gamma$ are at the end of the defined grid. This parameters give the shallowest density profile possible, still conserving the smoothness of the model. Further from that, the density profile stops being smooth and we encounter numerical errors when solving the radiative transfer equations in {\sc Turbospectrum}. The fact that the end of the grid gives the best possible fit is not surprising, since we are looking for the shallowest profile possible but still conserving a smooth profile. 

We notice that, although our best-fit model for HD~95687 can accurately reproduce both flux and $|V|^{2}$, this succeeds for V602~Car to a lower extent: the flux is well reproduced in Figure~\ref{fig:v602-car-2}, but the $|V|^{2}$ is still missing some extension, especially in the region $\lambda=2.29-2.5\,\mathrm{\mu m}$. As mentioned, this region not only includes CO but also the presence of water. A possible explanation for this mismatch could be that for increasing $\dot{M}$, the models still fail to reproduce the extension of the water or CO layers. Another possibility is that since our model neglects velocity gradients, it underestimates the equivalent widths of lines. Broader and stronger lines would help to increase the apparent stellar extension at those wavelengths. % As an additional explanation, since our model does not account for the velocity gradient, the presence of CO would be depleted causing shallower $|V|^{2}$ lines.   

\section{Summary and conclusion}
\label{sec:conclusion}
We present 1D modelling for the extended atmospheres of RSGs based on simple R.E. and chromospheric temperature inversion by \citet{2001ApJ...551.1073H}, and compute synthetic flux spectra and synthetic interferometric visibility spectra. When comparing our models to a simple {\sc MARCS} or {\sc PHOENIX} models, our synthetic $|V|^{2}$ shows a stronger atmospheric extension and can fit for the first time the observed extension in the case studies. 

%We have shown that the model can reproduce the spectra and visibilities convolved to any spectral resolution used by GRAVITY and MATISSE.
Regarding the temperature profile, we find that the R.E. reproduces the spectra better than the chromospheric temperature inversion, since we do not observe any emission in the CO bands, that are the result of models based on a temperature inversion. The possible reason that R.E. fits better than the temperature inversion, even though RSGs are known to have a chromosphere, could be: on the one hand, the presence of different spatial cells with different temperatures in the hot luke-warm chromospheres of RSGs \citep{2020A&A...638A..65O}. On the other hand, we do not know the effect that the dust could make where $T>800$ K, although we expect it to be small. 

Moreover, localised gaseous ejections, related to magnetic fields and surface activity were recently suggested as a major contributor to mass loss from RSGs \citep{2022AJ....163..103H,2022MNRAS.510..383A,2022A&A...661A..91L}. To explore this effect in detail, we would need to use 3D models, which is out of the scope of this paper. Our 1D modelling approach relies on an azimuthally averaged stratification, which is a good approximation for many aspects, but may not reproduce some of the observed features.

When compared to the observations, we obtain a mass-loss rate that is in accordance with typical mass-loss prescriptions \citep[e.g.,][]{1988A&AS...72..259D,2005ApJ...630L..73S,2020MNRAS.492.5994B}. However, in order to fit both the water and CO extensions simultaneously, the density shape should be steeper close to the surface of the star than previously expected by \citet{2001ApJ...551.1073H}.

Most importantly, we are able to reproduce the $|V|^{2}$ extension of the case studies. Simple stellar atmosphere models such as {\sc MARCS} do not show extension at all. However, the description very close to the stellar surface may not be optimum yet, as we are not able to reproduce some atomic lines in the $2.10\,\mathrm{\mu m}<\lambda<2.30\,\mathrm{\mu m}$ $|V|^{2}$ region, which are the most sensitive to the stratification very close to the stellar surface.

%\item Our model does not include the effect of dust in the extended atmosphere. 
  %%This result could be either due to the shape of the density profile, localized events in the stellar atmosphere \citep{2022AJ....163..103H,2022MNRAS.510..383A} or the need the include the dust at $>800$ K. For the first explanation, we have seen that changing the $\beta$-law shape can better fit the $|V|^{2}$ while keeping the SED untouched in the K-band. However, the TiO lines would change in the optical SED, moving the stars to later spectral types than expected. For the second and third hyphotesis, we would need to explore our models further including 3D components and dust modellization. 
%\item Overall, our model is able to reproduce both the $|V|^{2}$ and SED, and can open a window to study mass loss events of RSGs in the optical and near-IR without the need of dust modellization.
%\end{itemize}
\medskip
This is the first extended atmosphere model to our knowledge that can reproduce in great detail both the spectra and $|V|^{2}$ simultaneously. Therefore, we have shown the immense potential of this semi-empirical model of {\sc MARCS}+wind, not only to match the spectral features without the need of dusty shells, but also the visibilities obtained by interferometric means. 

In the future, we want to compare this model with data from a wider wavelength range, to see the full effects in higher wavelengths such as the $L$ or $M$-bands.   

\begin{acknowledgements}
We would like to thank the anonymous referee for their useful comments which helped to improve paper. Based on observations collected at the European Southern Observatory under ESO programme ID 091.D-0275. GGT is supported by an ESO studentship and a scholarship from the Liverpool John Moores University (LJMU). We want to thank A. Rosales-Guzman and J. Sanchez-Bermudez for useful discussions.

\end{acknowledgements}

% WARNING
%-------------------------------------------------------------------
% Please note that we have included the references to the file aa.dem in
% order to compile it, but we ask you to:
%
% - use BibTeX with the regular commands:
%   \bibliographystyle{aa} % style aa.bst
%   \bibliography{Yourfile} % your references Yourfile.bib
%
% - join the .bib files when you upload your source files
%-------------------------------------------------------------------
\bibliographystyle{aa}
\bibliography{sample}

\begin{appendix} %First appendix
\end{appendix}
\end{document}